\documentclass[apjl]{aastex}

\shorttitle{Limb-brightening in the restarted jet of 3C~84}
\shortauthors{Nagai et al.}

\begin{document}


\title{Limb-Brightened Jet of 3C~84 Revealed by the 43-GHz Very-Long-Baseline-Array Observation}


\author{H. Nagai\altaffilmark{1}, T. Haga\altaffilmark{2}, G. Giovannini\altaffilmark{3,4}, A. Doi\altaffilmark{5}, M. Orienti\altaffilmark {3,4}, F. D'Ammando\altaffilmark{3}, M. Kino\altaffilmark{5}, M. Nakamura\altaffilmark{6}, K. Asada\altaffilmark{6}, K. Hada\altaffilmark{7,3,1}, M. Giroletti\altaffilmark{3}}
\email{hiroshi.nagai@nao.ac.jp}



\altaffiltext{1}{National Astronomical Observatory of Japan, Osawa 2-21-1, Mitaka, Tokyo 181-8588, Japan }
\altaffiltext{2}{The Graduate University for Advanced Studies, 3-1-1 Yoshinodai, Chuo-ku, Sagamihara, Kanagawa 229-8510, Japan}
\altaffiltext{3}{INAF Istituto di Radioastronomia, via Gobetti 101, 40129, Bologna, Italy }
\altaffiltext{4}{Dipartimento di Fisica e Astronomia, Universita' di Bologna, via Ranzani 1, I-40127, Bologna, Italy }
\altaffiltext{5}{Institute of Space and Astronautical Science, Japan Aerospace Exploration Agency, 3-1-1 Yoshinodai, Chuo-ku, Sagamihara,\\ Kanagawa 229-8510, Japan }
\altaffiltext{6}{Institute of Astronomy \& Astrophysics, Academia Sinica, 11F of
Astronomy-Mathematics Building, AS/NTU No. 1, Taipei 10617, Taiwan}
\altaffiltext{7}{Research Fellow of the Japan Society for the Promotion of Science}

\begin{abstract}
We present a study of sub-pc scale radio structure of the radio galaxy 3C~84/NGC~1275 based on the Very Long Baseline Array (VLBA) data at 43~GHz.  We discover a limb-brightening in the ``restarted" jet associated with the 2005 radio outburst.  In the 1990s, the jet structure was ridge-brightening rather than limb-brightening, despite the observations being done with similar angular resolution.  This indicates that the transverse jet structure has changed recently.  This change in the morphology shows an interesting agreement with the $\gamma$-ray flux increase, i.e., the $\gamma$-ray flux in 1990s was at least seven times lower than the current one.  One plausible explanation for the limb-brightening is the velocity structure of the jet in the context of the stratified jet, which is a successful scenario to explain the $\gamma$-ray emission in some active galactic nuclei (AGNs).  If this is the case, the change in apparent transverse structure might be caused by the change in the transverse velocity structure.  We argue the possibility that the transition from ridge-brightening to limb-brightening is related to the $\gamma$-ray time variability on the timescale of decades.  We also discuss the collimation profile of the jet.
\end{abstract}	


\keywords{galaxies: active, galaxies: jets, galaxies: individual (3C~84, NGC~1275, Perseus~A), radio continuum: galaxies  
}	

\section{Introduction}
The radio source 3C~84 is associated with the giant elliptical/radio galaxy NGC~1275 (z=0.0176).  Its classification by radio luminosity is Fanaroff-Riley type-I \citep[e.g.,][]{Chiaberge1999}.  Thanks to its brightness and proximity, this source is one of the best-studied radio sources in history.  Recently increased activity starting in 2005 has been detected in radio band \citep{Abdo2009}.  The Very Long Baseline Interferometry (VLBI) observations revealed that this flux density increase originated within the central pc-scale core, accompanying the ejection of a new jet component \citep[hereafter Paper I]{Nagai2010}.  This new component appeared from the south of the core around 2003, and is moving to the position angle $\sim160^{\circ}$ steadily with slightly changing speed in both parallel and perpendicular directions \citep[hereafter Paper II]{Suzuki2012}.  The apparent speed ranges from $0.1c$ to $0.5c$ (Paper II).  The moving direction of the new component is clearly different from the direction of the pre-existing component near the core separated about 0.5~mas along the position angle of $-150^{\circ}$.  The flux density of both the new component and the core particularly increased after 2007-2008, and the new component has shown a further increase in flux density after 2009\citep[hereafter Paper III]{Nagai2012}.

It is notable that 3C~84 is the best studied $\gamma$-ray radio galaxy along with M~87 and Centaurus~A, and therefore these sources are the ideal laboratories to study the $\gamma$-ray emission mechanism in misaligned Active Galactic Nuclei (AGNs).  Strong variability in the $\gamma$-ray emission has been detected from 3C~84 by the Large Area Telescope (LAT) on board {\em Fermi}.  An averaged $\gamma$-ray flux during the first four months is $(2.10\pm0.23)\times10^{-7}$~ph cm$^{-2}$ s$^{-1}$ above 100~MeV.  This $\gamma$-ray flux is seven times brighter than the upper limit estimated by EGRET/{\em CGRO}.  It was claimed that the innermost jet of 3C~84 is the most likely source of $\gamma$-ray emission because of the different $\gamma$-ray activity level between EGRET era and {\it Fermi} era \citep{Abdo2009}.  During the first two years of observation, variations on the timescale of a month were observed by {\em Fermi}-LAT in $\gamma$-rays: one occurred in 2009 April-May \citep{Kataoka2010}, and the other one occurred in 2010 June-August \citep{BrownAdams2011}.  In particular, during the second flare photons up to 102.5 GeV were detected.  The largest increased activity in GeV band was reported on 2013 January 21 \citep{Ciprini2013}.  An average daily $\gamma$-ray flux is six times greater than the average flux reported in the second {\em Fermi}-LAT catalog. 

In Paper III, the radio time variability was studied to search for possible correlation with the $\gamma$-ray variability, but no clear correlation was found on the timescale of $\gamma$-ray time variation.  Neither new component ejection nor change in morphology associated with the $\gamma$-ray flares were found by VLBI observations (Paper III).  The lack of significant changes in radio band for 3C~84 after the detection of high $\gamma$-ray activity leaves the debate on the region responsible for the high-energy emission and its location still open.  Besides, the apparent speed detected by VLBI is relatively slower than the jet speed predicted from one-zone synchrotron-self Compton model or deceleration jet model unless the jet angle to the line of sight is very small ($<5^{\circ}$: Paper II).  This indicates that the gamma-ray emitting region may have a higher Lorentz factor, and that the emission may be more strongly beamed than would be implied by the Lorentz factor estimated from the VLBI proper motions.  We also performed spectral energy distribution (SED) fit to the observed broadband spectrum using the estimated apparent speed, but it failed to reproduce the optically thin radio spectrum observed by VLBI(Paper II).

While several possibilities to reconcile the discrepancy between radio and $\gamma$-ray properties have been discussed in earlier works (Paper I; Paper II; Paper III), it is of great interest to investigate the presence of stratified structure with a velocity gradient \citep{Ghisellini2005} in the jet of 3C~84.  
According to this scenario, the radio emission is mostly coming from the slower sheath while the emission from the spine is beaming away from the line of sight.   Therefore, the limb-brightening can be observed along the jet if the spine-sheath structure is present.  Clear evidence of limb-brightening is found in several AGNs such as M~87\citep{Junor1999}, Mrk~501\citep{Giroletti2008}, Mrk~421\citep{Piner2010}, and 1144+35\citep{Giovannini1999}, but no clear signature of limb-brightening has been found in 3C~84 so far.  To detect the limb-brightening, high spatial resolutions that can resolve the transverse direction of the jet and high dynamic range image are required.  In this paper, we report a new VLBA observation at 43~GHz to investigate the transverse structure of the jet in 3C~84.  Mainly we focus on discussing the origin of limb-brightening, but we also argue the collimation profile that can be obtained by resolving transverse direction.

At the 3C 84 distance 1~mas corresponds to 0.344~pc assuming $H_{0}$=70.5~km~s$^{-1}$~Mpc$^{-1}$, $\Omega_{\rm M}=0.27$, and $\Omega_{\Lambda}=0.73$.  Assuming a super massive black hole (SMBH) mass of $3.4\times10^{8}~M_{\bigodot}$ \citep{Wilman2005}, $10^{3}$ Schwarzschild radii ($r_{g}$) corresponds $9.5\times10^{-2}$~mas. 
	
\section{Observations and Data Analysis}
The observation was carried out with 10 VLBA stations at 43~GHz on 2013 January 24.  The data consists of eight IFs with a 32-MHz bandwidth for each IF.  Total bandwidth is 256~MHz per polarization.  Both right-hand and left-hand circular polarizations were obtained, and only parallel-hand correlations (RR and LL) were obtained in the correlation process because the polarization information is out of the scope of this observation.  We performed phase-referencing mode observation by switching the pointing between the target source and the calibrator J0313+4120.  The observation consisted of many different scans for 3C~84 and the calibrator as well.  The integration time for each scan is ten seconds.  Total observing time for 3C~84 is about 45 minutes, and overall observation including calibrator scans were spanned over eight hours.  The scans for 3C~84 were spread over different hour angles evenly, and therefore we obtained good $uv$-coverage.  In this paper, we do not present phase-referencing image but focus on the self-calibrated image.  The purpose of phase-referencing analysis is to measure the core-shift effect \citep[e.g.,][]{Hada2011}.  This will be reported in a forthcoming paper (Haga et al. in preparation).

The data reduction was performed using Astronomical Imaging Processing System (AIPS) developed by National Radio Astronomy Observatory (NRAO).  An {\it a priori} amplitude calibration was performed using aperture efficiency and system noise temperature provide by each station.  The opacity correction for the atmospheric attenuation was applied.  Fringe fitting and bandpass calibration were applied.  For the deconvolution of synthesized image, we used CLEAN and self-calibration technique. Final images were obtained after a number of iterations with CLEAN and both phase and amplitude self-calibration using the DIFMAP software package \citep{Shepherd1994}.
 
\section{Results}
\begin{figure}
\includegraphics[width=8.5cm]{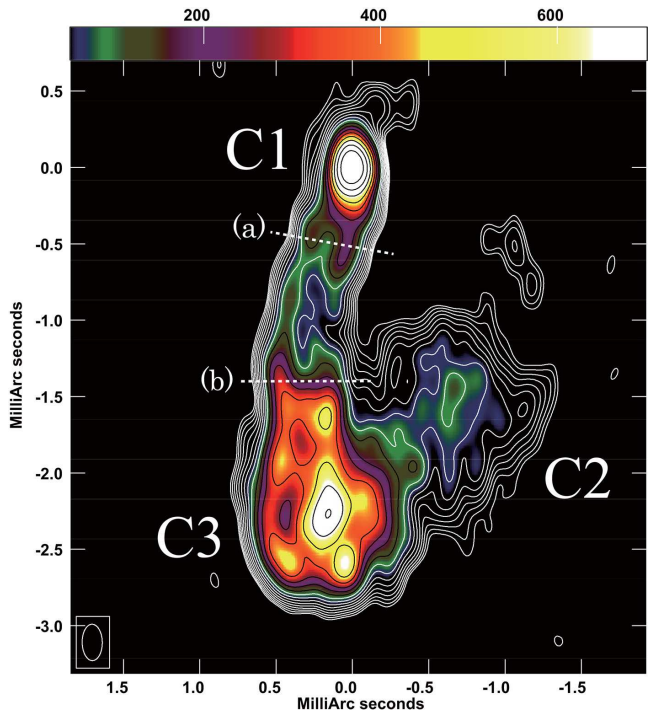}
\caption{The 43-GHz total intensity map of 3C~84.  The contours are plotted at the level of 5.43$\times$($-$1.41, 1, 1.41, 2.83, 4, 5.66, 8, 11.3, 16, 22.6, 32, 45.3, 64, 90.5, 128, 181, 256)~mJy~beam$^{-1}$.  The peak intensity is 2.17~Jy~beam$^{-1}$.  The color range is from 100~mJy~beam$^{-1}$ to 700~mJy~beam$^{-1}$.  The ellipse shown at the bottom left corner of the image indicates the full-width half-maximum of the convolved beam.  The FWHM of the convolved beam is $0.24\times0.13$~mas at the position angle $0.69^{\circ}$.  The white broken lines (a) and (b) indicated in the figure is the slice locations for Fig. \ref{slice}.}
\label{Map}
\end{figure}

Figure \ref{Map} shows the self-calibrated image of 3C~84.  In previous studies it was shown that the pc-scale structure mostly consisted of three components.  We have detected the same structure, but a finer scale structure is visible from our image.  The bright core and one-sided jet structure is clearly seen.  The jet position angle is $-170^{\circ}$ up to 1.2~mas from the core and then slightly changes to the position angle $-180^{\circ}$.  At the end of the jet, there is a bright knot-like feature.  While this feature was represented by a single Gaussian component (labeled C3 in Figure \ref{Map}) in the previous studies, multiple subcomponents are seen from this image.  The changing pattern of the jet direction is approximately consistent with the previously detected path of C3 motion (Paper II).  There is an elongated feature (C2) towards the west from C3 region, which invokes the backflow from C3.  However, C2 had already been present before the emergence of C3.  Therefore, the origin of C2 and its connection with C3 is not very clear.  We did not detect any significant emission from the counter-jet side.

\begin{figure}
\includegraphics[width=7.5cm]{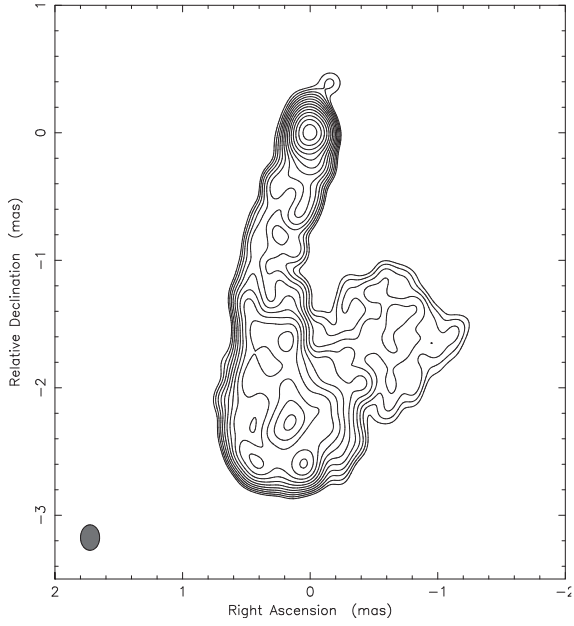}
\caption{Same map with Fig. \ref{Map}, but the contours are plotted at the -20, 20, 28, 40, . . ., 640~mJy~beam$^{-1}$ with the convolved beam FWHM 0.20 $\times$ 0.15~mas at the position angle 0 degree, which are the same contour levels and beam size as those reported in \cite{Dhawan1998}.}
\label{Map_DhawanLevel}
\end{figure}

The most remarkable finding is that the limb-brightening is evident along the approaching jet, which is the ``restarted" jet associated with an ongoing activity started from 2005.  Similar quality of images have been available from the website of Boston University's blazar monitoring program\footnote{http:\slash\slash{}www.bu.edu\slash{}blazars\slash{}VLBAproject.html} since 2010 November 1.  At that time the morphology of the 3C~84 jet was limb-brightened, and all images available up to 2013 July 28 are in very good agreement with the image presented here.  In the VLBA 43-GHz images during the period 2002-2008 (Paper II), the transverse structure was not very clear.  This is probably due to the lack of dynamic range.  \cite{Dhawan1998}, \cite{Romney1995}, and \cite{Lister2001} reported the 43-GHz VLBA images of 3C~84 as of 1990s.  From those images, no clear limb-brightening was seen despite similar angular resolution was achieved.  One might think that this is also due to the lack of dynamic range.  However, we note that the most sensitive observation by \cite{Dhawan1998} achieved better image noise than our observation, thanks to much longer integration time  (73\% duty cycle for 14 hours) and participation of 1 VLA antenna.  We also created the image with the same contour levels and convolved beam as presented in \cite{Dhawan1998}, but the limb-brightening is still visible (see Figure \ref{Map_DhawanLevel}).  Therefore, the apparent transverse structure of the jet has indeed changed recently, and this change has occurred at least before 2010 November (first epoch of Boston program) and after 1999 April (the observation epoch of \cite{Lister2001}).  

\begin{figure}
\includegraphics[width=7.5cm]{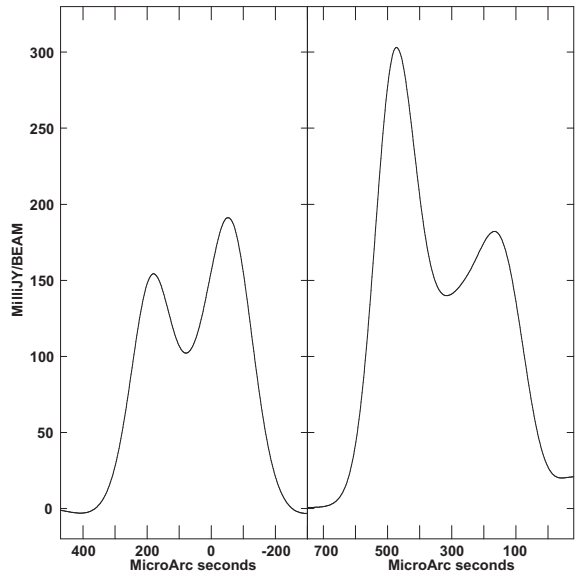}
\caption{The transverse slice profile of the jet.  The left and right figures show the profiles along (a) and (b) indicated by the broken lines in Fig. \ref{Map}, respectively. }
\label{slice}
\end{figure}

\begin{figure}
\includegraphics[width=8cm]{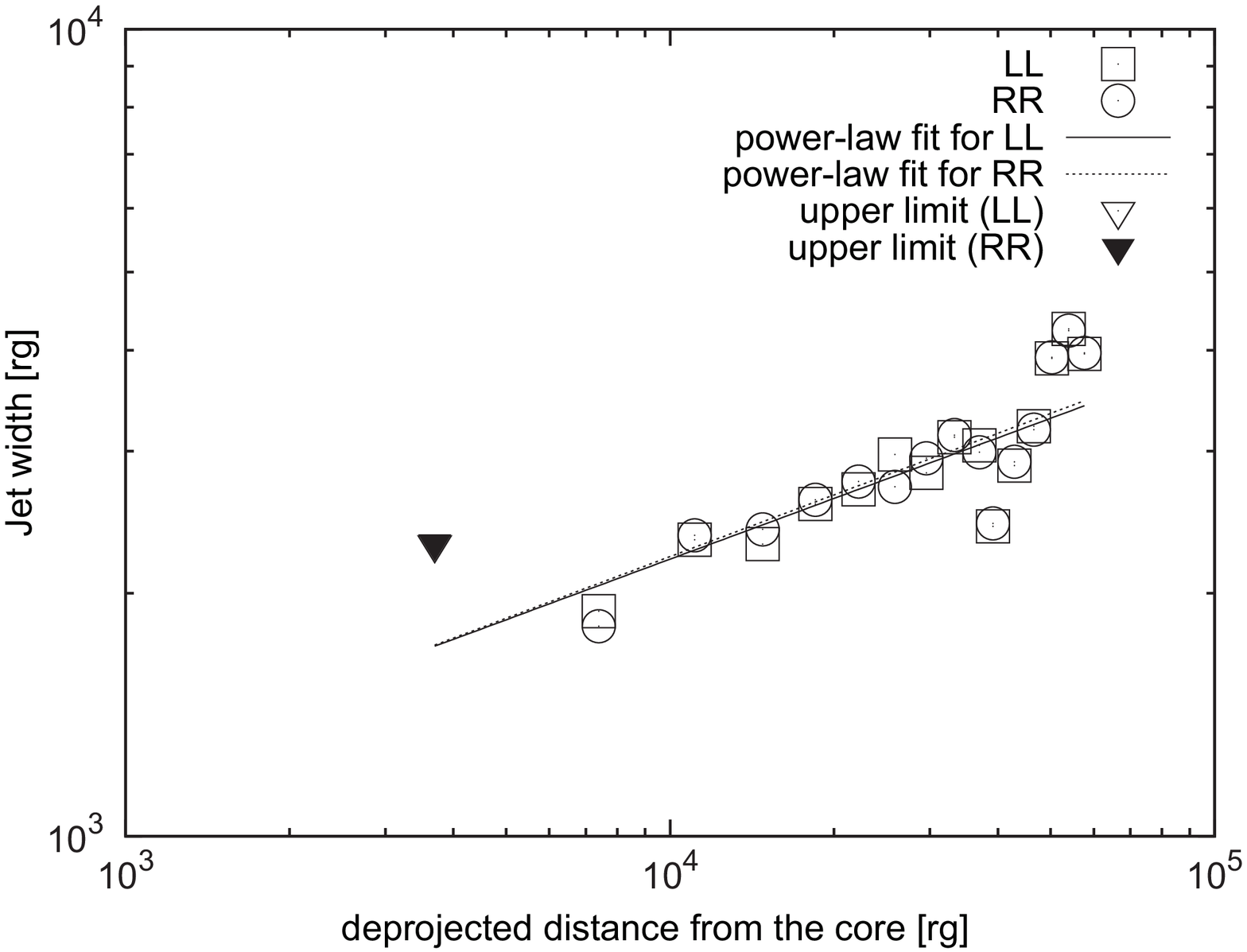}
\caption{The jet width profile.  The open squares and open circles represent the measurement for LL and RR maps, respectively.  The filled triangle indicates the upper limit constrained by the FWHM of a single Gaussian component.  The error of jet width is smaller than the size of each symbol.  The horizontal axis is shown in the deprojected distance assuming with the viewing angle of $25^{\circ}$.  Both vertical and horizontal axes are shown in the unit of Schwarzschild radius.  Here we adopt $3.4\times10^{8}~M_{\bigodot}$ for the black hole mass \citep{Wilman2005}.  The broken and dotted lines are the power-law fits for LL and RR data, and the resultant power-law indices are $0.25\pm0.03$ for both cases.}
\label{JetWidth}
\end{figure}

Resolved jet in transverse direction allows us to study the jet width profile. We produced the slice profiles of the total intensity across the jet at different cross-sections along the jet.  The slices were made at every 0.15~mas step (roughly the same size as the synthesized beam) from the core along the position angle $-10^{\circ}$ up to 1.5~mas and along the position angle $0^{\circ}$ beyond.  Examples of the slice profiles are shown in Figure \ref{slice}.  The slice profiles are well represented by two Gaussian components.  To evaluate the jet width, we fitted two Gaussians to the slice profiles and measured the separation between the peaks of the Gaussians.  In some cases an additional Gaussian component between two Gaussians was required, and we regarded the separation between two outer components as the jet width.  The slice at 0.15~mas from the core is represented by a single Gaussian due to the lack of resolution.  We regard the FWHM as the upper limit of the jet width.  Figure \ref{JetWidth} shows the jet width as a function of the distance from the core.  The power-law fit yields a power-law index of $0.25\pm0.03$.   The change in the jet width as a function of the distance indicates the collimation profile of the jet, which is a key quantity to study the jet formation mechanism.  This will be discussed in section \ref{sect4.2}.  The issue of how to determine the jet width may be controversial.  We also evaluated the jet width by the separation between the outer sides of the half-maximum points of the two Gaussians.  The resultant power-law index is $\sim0.31$, which shows no significant difference from the case that the width is defined by the peaks of the two Gaussians.  In this paper, we adopt the former case for the definition of the jet width.

For the consistency check, we performed the calibration and subsequent imaging for RR and LL correlations independently.  The Gaussian fit to both RR and LL images shows a good agreement (see Figure \ref{JetWidth}).  

\section{DISCUSSION}

\subsection{Limb-brightening}
The detection of limb-brightening is expected as if there is a velocity gradient across the jet and the beaming-cone angle of the emission from the ``spine" is smaller than the jet viewing angle \citep[e.g.,][]{Giroletti2004, Kataoka2006}.  So far we have not detected a clear correlation in light curve between radio and $\gamma$-ray in 3C~84 on the timescale of days to months.  Also, the observed apparent motion is relatively slower than the jet velocities expected from the $\gamma$-ray emission models.  Yet, such a lack of correlation between radio and $\gamma$-ray is expected as the radio emission is dominated by the synchrotron radiation from the slow sheath and the $\gamma$-ray emission is dominated by the Comptonization of both the spine and sheath photons by the electrons in the sheath (see bottom panel of Figure 5 in \cite{Ghisellini2005}).  Such a ``spine-sheath" structure is also favored to explain the $\gamma$-ray luminosity.  The observed $\gamma$-ray luminosity of 3C~84 by {\it Fermi}-LAT is of the order of $10^{44}$~erg~s$^{-1}$.  This is already comparable to the typical observed $\gamma$-ray luminosity of low-frequency peaked BL~Lac objects (LBL), which are the small-viewing angle counterparts of FRI radio galaxies in the context of unification scheme \citep{Urry1995}, despite the Doppler factor of radio galaxies would be smaller than that of LBLs.  This problem would be eased if the 3C~84 is dominated not by the fast spine but by the slow sheath \citep{Kataoka2010}.

Here we roughly estimate the outflow velocity of the limb-brightened region.  For simplicity, we assume an intrinsically uniform brightness across the jet in the transverse direction and the observed transverse brightness is only affected by the transverse velocity structure that results in the varying Doppler enhancement of different jet axisymmetric layers.  With this assumption, the limb-brightened structure corresponds to a layer where the Doppler factor becomes maximum.  Assuming the viewing angle ($\theta$) of $25^{\circ}$ adopted in \cite{Abdo2009}, the Doppler factor reaches its maximum where the bulk Lorentz factor ($\Gamma$) is about 2.4 ($\beta\sim0.9$).  Thus for having a limb-brightening morphology, as observed in 3C~84, we require the velocity of the limb-brightened region (sheath) and the inner dim region (spine) to be $\Gamma\sim2.4$ and $\Gamma\gg2.4$, respectively.  Estimated Lorentz factor of limb-brightened region seems somewhat faster than the VLBI-measured velocity of C3 in Paper I and II.  This may indicate that C3 is the terminal hotspot and its motion does not reflect the jet flow itself.  

Apparently, the resultant velocity constraints depend on the viewing angle.  In principle, it is possible to give a constraint on the viewing angle from the jet/counter-jet intensity ratio argument.  However, in the case of 3C~84, non-detection of counter-jet is not only due to the Doppler effect but also due to the free-free absorption effect as reported in \cite{Walker2000}.  This makes it difficult to obtain a robust constraint on the viewing angle.  The detection of counter jet by higher dynamic range observation would be useful to get a constraint on the viewing angle, providing us an indication if this value changed respect to previous estimations \citep{Walker2000, Asada2006, Lister2009}.  

\cite{Dhawan1998}, \cite{Romney1995}, and \cite{Lister2001} reported the 43-GHz VLBA images of 3C~84 as of 1990s.  From those images, no clear limb-brightening was seen despite similar angular resolution was achieved.  One possibility to explain this difference with assuming a constant jet viewing angle and intrinsically uniform brightness distribution across the jet is a change in the transverse velocity structure.  If the flow velocity in the spine has become faster, such a change can be produced.  

Alternatively, the apparent change of jet structure from ridge-brightening to limb-brightening could be explained by a change in the jet viewing angle with constant jet physical properties.  If the jet viewing angle was very small in the 1990s while the recent viewing angle is moderate, the ridge-brightening could appear only in the 1990s because of strong beaming of the spine emission towards the observer.  
However, we feel that this possibility is very unlikely.  If the viewing angle was small and the physical properties of the jet (such as magnetic field, electron energy distribution) were similar to the current ones, 3C~84 should have been detected as a bright $\gamma$-ray source because of the strong beaming effect.  Contrary to this, the upper limit of $\gamma$-ray flux estimated by EGRET in 1990s was at least seven times weaker than the recent flux level.  

An interesting agreement is that the ridge-brightening appeared in a low $\gamma$-ray state and the limb-brightening appears in an on-going high $\gamma$-ray state.  In the spine-sheath structure, the external Compton in the sheath is the dominant contributor to $\gamma$-ray emission at large viewing angles \citep{Ghisellini2005, Tavecchio2008}.  Thus, we speculate that the origin of the $\gamma$-ray high state might be the result of an increase in the energy of the seed photons from the spine related to a possible increase of the spine flow velocity as well as in a brightness increase of the sheath region. 
How such a change in the transverse velocity can occur is an open question, but the restarted activity of 3C~84 may be tied to this change. Yet, the sheath Doppler factor of 2.4 ($\delta\sim\Gamma\sim2.4$, where $\delta$ is the Doppler factor) could be problematic for the synchrotron peak position.  Recent SED modelings including very high energy (VHE) $\gamma$-ray data in 2009 October - 2010 February and 2010 August - 2011 February shows that $\delta=2$ shift the synchrotron bump to lower frequencies and the resultant SED disagree with the observed optical, UV, and X-ray data. Higher Doppler factor ($\delta\sim4$) is favored to reproduce those data \citep{Aleksic2013}.  However, we note that this SED modeling does not fit to the radio data.  This may require an additional emitting component to reproduce the whole electromagnetic spectrum.  In any case, the limb-brightened structure in 3C~84 jet suggests that the jet property in transverse direction is not uniform and this should be taken into account for the broadband SED modeling.
   
The transverse velocity structure is just one possibility to explain the limb-brightening and we cannot exclude other possibilities.  Here let us briefly comment on them.  The limb-brightening can result from the helically wrapped magnetic field structure as discussed to explain the limb-brightening of M~87 \citep{Owen1989}.  More recently, \cite{Clausen-Brown2011} analyzed in detail the transverse intensity profile of a cylindrical jet with helical magnetic field for various conditions.  It is shown that a limb-brightening is evident in the case of $\theta_{\rm ob}\sim1$~rad, where $\theta_{\rm ob}$ is the jet viewing angle in the observer frame.  It should be noted that the observed brightness ratio between eastern and western limbs in 3C~84 is not constant.  The western limb is brighter near the core ($\sim0.5$~mas from the core) while the eastern limb is brighter between $\sim0.5$~mas and $\sim1.5$~mas from the core (see Figure \ref{Map}).  Then, the western limb becomes brighter again beyond $\sim1.5$~mas from the core.  This side-to-side change of brightness pattern is similar to a filament cutting diagonally across the jet, invoking the helically wrapped magnetic field.   However, it seems to be difficult with this model to explain the change from ridge-brightening to limb-brightening unless the helical magnetic field is present only in the restarted jet.  It is also possible to reproduce the limb-brightening if the particle acceleration is more efficient near the jet boundary \citep{Ostrowski1990, Stawarz2002, Rieger2004}.  With this framework, the coincidence of the $\gamma$-ray time variability and the apparent structural change invokes that the Comptonization of the synchrotron emission from the boundary shear is an origin of the $\gamma$-ray emission.  But, this model is not favorable to explain the non-detection of the radio counterpart of $\gamma$-ray flares (Paper III).  \cite{Zakamska2008} also reported that a pileup of material along the jet boundary is produced by the interaction between the jet and the ambient medium when the jet has a high Lorentz factor.  They claimed that such a boundary pileup may be the reason for the limb-brightening.  Yet, the observed apparent velocity in sub-pc region in 3C~84 is sub-relativistic, which is not compatible with this condition.

\subsection{Jet width}\label{sect4.2}
The change in jet width with the distance probes the collimation profile, which is a key parameter to discriminate the theoretical models of jet formation.  Theoretical study suggests that the external pressure is essential to determine the profile of the global collimation.  \cite{Komissarov2009} show that when the external gas pressure follows $p_{\rm ext}\propto r^{-a}$ where $a<2$, the jet maintains a parabolic shape ($W_{j}\propto r^{a/4}$, where $W_{j}$ is the jet width and $r$ is the distance from the core), whereas for $a>2$ the jet eventually becomes conical due to insufficient external support.  As shown in section 3, the observed jet width as a function of the distance for 3C~84 is given by $W_{j}\propto r^{0.25\pm0.03}$.  If the limb-brightening region represents the exterior of the magnetized jet, our observation indicates $a=1$.  Similar studies were done for the jet of M~87, and the collimation profile showed rather parabolic shape ($W_{j}\propto r^{0.56}$) in $r_{g}>10^{4}$ \citep{Asada2012, Hada2013}.  The apparent rapid collimation in 3C~84 could indicate that the ambient pressure gradient is shallower than that in M~87.  No rapid broadening of the jet width was reported in another radio galaxy Centaurus~A \citep{Muller2011}.  

Since M~87 and Centaurus~A are resolved in a deeper inner region, it is important to study the inner-jet collimation property of 3C~84 with higher angular resolution from the viewpoint of comparison.  In particular, a possible change in the collimation profile from the outer one, which is reported in $<100~r_{g}$ scale in M~87 \citep{Hada2013}, is of great interest.  So far this region has only been accessible in M~87.  (Sub)mm VLBI observations will probe such an inner region in other radio galaxies including 3C~84, thereby allowing a more comprehensive understanding of jet formation.  It is also interesting to see how the collimation profile changes as we go further downstream along the jet.  One might expect that the jet radius expands more rapidly when the jet goes out from the dense environment \citep{Komissarov2009}.  In 3C~84, the restarted jet is still expanding, and therefore monitoring the jet transverse structure will allow us to study the collimation profile along the jet at larger distance from the core.

\section{Conclusion}
Clear limb-brightening is discovered in the sub-pc scale jet of 3C~84, which has been formed by the restarted activity observed since 2005, while the jet structure was rather ridge-brightening in the 1990s.  Although several possibilities are considered, we propose that the change from ridge-brightening to limb-brightening may be attributed to a change in the transverse velocity structure on the basis of a ``spine-sheath" scenario.  This is compatible with the lack of the radio counterpart of short-term $\gamma$-ray flares reported in Paper II.  The jet width profile in $\sim10^{3}$-$10^{5}$~$r_{g}$ scale is rapidly collimated rather than parabolic shape and differs from the trend in M~87, which might reflect the different circumnuclear environment between M~87 and 3C~84.  Higher resolution studies are important to probe whether the limb-brightening is present in the innermost region of 3C~84 and study the collimation profile as well.  Future high-resolution instruments, such as (sub)mm VLBI including phase-up ALMA, will give us a hint for the jet physics in connection with the $\gamma$-ray emission and jet formation theory.  

\acknowledgments
We thank an anonymous referee for constructive comments to improve the manuscript.  We also thank F. Tavecchio and K. Akiyama for their helpful comments on the spine-sheath model and interpretation.  Part of this work was done with the contribution of the Italian Ministry of Foreign Affairs and University and Research for the collaboration project between Italy and Japan.  This work is partially supported by the Grant-in-Aid for Scientific Research, KAKENHI 24540240 from the Japan Society for the Promotion of Science (JSPS).  The VLBA is operated by the US National Radio Astronomy Observatory (NRAO), a facility of the National Science Foundation operated under cooperative agreement by Associated Universities, Inc.



\renewcommand{\bibname}{}

\end{document}